\documentclass[journal=aamick,manuscript=article]{achemso}
\usepackage[colorlinks=true, citecolor=blue]{hyperref}
\newcommand{\AK}[1]{\textcolor{black}{#1}}

\usepackage[version=3]{mhchem} 

\author{Asish K. Kundu}
 \affiliation{National Synchrotron Light Source II, Brookhaven National Laboratory, Upton, New York 11973, USA}
\email{akundu@bnl.gov}
 \author{Anil Rajapitamahuni}
  \affiliation{National Synchrotron Light Source II, Brookhaven National Laboratory, Upton, New York 11973, USA}
  \alsoaffiliation{Department of Physics, SRM University -- AP, Amaravati, Andhra Pradesh, 522502,India}
  \alsoaffiliation{Department of Physics and Astronomy, University of Nebraska-Lincoln, Nebraska 68588, USA}
\author{Niraj Aryal }
    \affiliation{Condensed Matter Physics and Materials Science Division, Brookhaven National Laboratory, Upton, New York 11973, USA}
  \author{Turgut Yilmaz}
   \affiliation{Department of Physics, University of Connecticut, Storrs, CT 06269, USA}
   \alsoaffiliation{Department of Physics, Xiamen University Malaysia, Sepang 43900, Malaysia}
 \author{ Margalit L. Feuer}
   \affiliation{Columbia University, Department of Chemistry, New York, NY, USA}
  \author{Suji Park}
   \affiliation{ Center for Functional Nanomaterials, Brookhaven National Laboratory, Upton, NY, USA}
  \author{Houk Jang}
     \affiliation{ Center for Functional Nanomaterials, Brookhaven National Laboratory, Upton, NY, USA}
  \author{Abhay Pasupathy}
   \affiliation{Condensed Matter Physics and Materials Science Division, Brookhaven National Laboratory, Upton, New York 11973, USA}
  \alsoaffiliation{Columbia University, Department of Physics, New York, NY, USA}
  \author{Xavier Roy}
  \affiliation{Columbia University, Department of Chemistry, New York, NY, USA}
 
 \author{Jerzy T. Sadowski}
   \affiliation{ Center for Functional Nanomaterials, Brookhaven National Laboratory, Upton, NY, USA}
\author{Elio Vescovo}
 \affiliation{National Synchrotron Light Source II, Brookhaven National Laboratory, Upton, New York 11973, USA}

\title[An \textsf{achemso} demo]
  {Proximity-induced charge transfer, strain and magnetic exchange in graphene/CrSBr heterostructure}

\begin{document}


\begin{abstract}
  Stacking van der Waals materials provides a powerful route to engineer emergent electronic and magnetic behaviors through proximity-driven interactions. The graphene/CrSBr heterostructure has emerged as a compelling platform in this frontier, exhibiting exotic macroscopic responses--including uniaxial surface-plasmon-polariton propagation and an unconventional quantum Hall effect-- indicative of strong interfacial electronic and magnetic coupling. However, a microscopic electronic landscape governing these phenomena has remained elusive. Here, we provide a comprehensive spectroscopic characterization of the graphene/CrSBr interface using a combination of angle-resolved photoemission spectroscopy (ARPES), low-energy electron microscopy (LEEM), and density functional theory. We resolve a massive redistribution of interfacial charge that concurrently hole dopes graphene and populates the quasi-one-dimensional spin-polarized conduction band of CrSBr, resulting insulator-to-metal transition in the interfacial CrSBr layer. Furthermore, electronic structure of CrSBr exhibits strong momentum‑dependent renormalization distinct from conventional charge doping, and theoretical modeling supported by Raman spectroscopy points to additional interfacial compressive strain. In addition, ARPES reveals a splitting-like feature in the graphene Dirac cone consistent with spin degeneracy lifting, providing possible evidence of magnetic proximity coupling. These findings provide crucial microscopic insight into the system’s optical and transport responses and establish graphene/CrSBr as a versatile platform for charge‑transfer control, strain‑driven band engineering, magnetic‑proximity coupling, and directionally confined excitations for next‑generation spintronic and nanophotonic devices.
\end{abstract}

\section{Introduction}
Van der Waals (vdW) heterostructures provide a versatile platform for engineering electronic, magnetic, and optical functionalities that do not exist in their constituent layers. By stacking atomically thin crystals with distinct symmetries, interaction strengths, and electronic structures, it is possible to generate emergent states driven by interfacial charge transfer, proximity coupling, and lattice strain. These interfacial mechanisms underpin a wide range of collective phenomena, including moiré flat bands, correlated insulating phases, unconventional superconductivity, proximity‑induced magnetism, and highly anisotropic plasmonic responses—and continue to motivate the search for new material combinations with tunable quantum properties \cite{lisi2021observation,cao2018unconventional,rizzo2025engineering,yang2024electrostatically,rossi2023direct,xia2025superconductivity,zhao2026second,li2025stacking,feuer2025charge}. In particular, interfacing graphene (Gr) with two-dimensional (2D) magnetic semiconductors/insulators offers a pathway to ``imprint" magnetic order onto high-mobility Dirac fermions, potentially realizing spin-polarized states and proximity-induced exchange interactions essential for next-generation spintronics \cite{yang2024electrostatically,rassekh2026proximity,voloshina2026substrate,hallal2017tailoring,yang2013proximity}.

A compelling candidate for such interfacial engineering is CrSBr, an air-stable, layered A-type antiferromagnetic (AFM) insulator with a Neel temperature of $T_N \approx 132$ K) \cite{wilson2021interlayer,telford2022coupling,ziebel2024crsbr}, nearly twice that of the widely studied CrI$_3$ and CrGeTe$_3$ systems \cite{liu2018anisotropic,kundu2020valence}. Its magnetic structure is defined by robust intralayer ferromagnetic exchange and antiferromagnetic coupling between adjacent layers \cite{telford2022coupling} and electronic structure hosts quasi‑one‑dimensional (1D) spin‑polarized states near the Fermi level \cite{Watson_2024, smolenski2025large,bianchi2023charge}, making it an intriguing platform for studying strongly correlated phenomena. Although its tunable layer- and doping-dependent magnetic order \cite{telford2022coupling,zhao2025doping,feuer2025charge}, interplay between magnetism and optical emission \cite{marques2023interplay}, and spin-polarized bands \cite{Watson_2024,klein2022control} make it a premier candidate for optoelectronic and spintronic devices, recent investigations suggest even deeper synergies when integrated into vdW heterostructures.

In particular, Gr/CrSBr heterostructures host a rich set of interfacial phenomena, including uniaxial propagation of surface plasmon‑polaritons (SPPs) in graphene \cite{rizzo2025engineering}, strong spin-charge coupling \cite{ghiasi2021electrical}, and an unconventional quantum Hall effect \cite{yang2024electrostatically}. The unidirectional SPPs are attributed to interfacial charge transfer and occupation of CrSBr’s quasi‑1D conduction band, providing a new route to breaking graphene’s otherwise isotropic plasmonic response \cite{huidobro2016graphene}. Such direction‑selective plasmon transport enables low‑loss, waveguide‑like energy flow and deterministic light-matter coupling, both essential for next-generation nanophotonics. In parallel, the unconventional quantum Hall effect points to counter-propagating, spin-polarized edge channels \cite{ghiasi2021electrical}, hypothesized to originate from a proximity-induced exchange shift in the graphene $\pi$-bands that imprints CrSBr’s magnetic topology onto the Dirac fermions.

However, despite these intriguing optical and transport signatures, a direct, momentum-resolved visualization of the electronic reconstruction at the Gr/CrSBr interface and the microscopic origin of these phenomena, is fundamentally missing. Several key questions remain regarding the interfacial band alignment, the magnitude of charge transfer, the role of interfacial strain on the intrinsic bandgap of CrSBr and whether the magnetic proximity is robust enough to induce a measurable exchange splitting in the graphene Dirac cone. \AK{Furthermore, understanding how many-body interactions renormalize the electronic structure at the Gr/CrSBr interface is essential for interpreting the electronic reconstruction and its coupling to magnetism.}

In this work, we address these questions by providing a comprehensive spectroscopic mapping of the bulk CrSBr and Gr/CrSBr heterostructure using angle-resolved photoemission spectroscopy (ARPES), and low-energy electron microscopy (LEEM), supported by density functional theory (DFT) calculations. We demonstrate substantial interfacial charge transfer that heavily hole dopes graphene and partially occupies the 1D-like CrSBr conduction band. This redistribution drives a pronounced renormalization of the CrSBr electronic structure and an insulator-to-metal transition in Gr/CrSBr. By comparing experiment and theory, we attribute this band renormalization to the combined effects of interfacial charge transfer and compressive strain in the interfacial CrSBr layer. \AK{Finally, we report splitting-like feature in the graphene $\pi$ band, which appears consistent with a possible spin-degeneracy lifting induced by magnetic proximity. More broadly, our findings highlight how interfacial proximity effects can be harnessed to engineer and tailor quantum states in two‑dimensional heterostructures.}

\section{Experimental results and discussions}

\begin{figure}[h!]
\centering
\includegraphics[width=8cm]{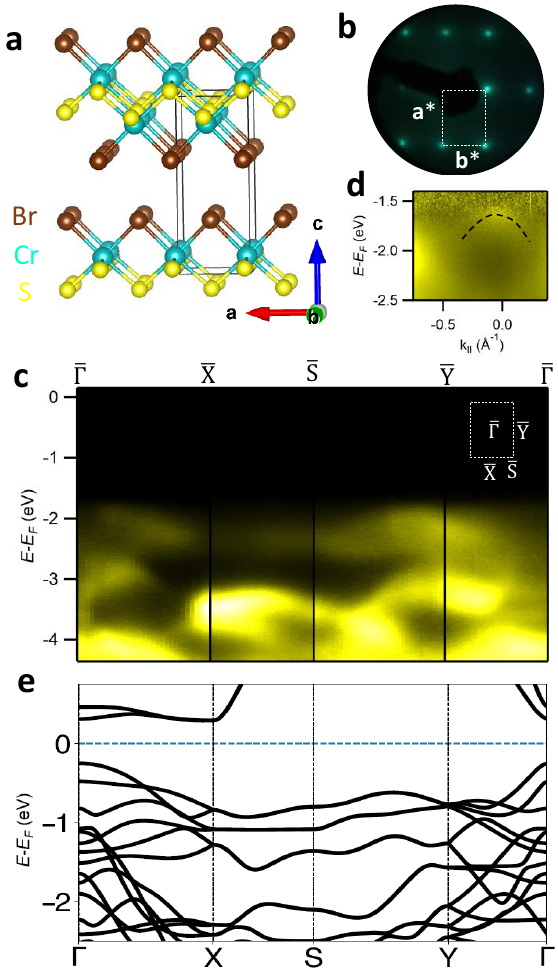}
\caption {Structure and electronic structure of CrSBr. (a) Crystal structure of CrSBr. (b) Low-energy electron diffraction pattern from a cleaved CrSBr surface. (c) ARPES spectra along high symmetry path measured with 21.2 eV photons at room temperature. (d) Zoomed-view of the top part of the valence band around $\bar{\Gamma}$. Parabolic dispersion of the valence band is highlighted by dashed curve. The bright and dark colors correspond to the maximum and minimum photoemission intensities, respectively. (e) Electronic structure of bulk CrSBr along $\Gamma$-X-S-Y-$\Gamma$ from DFT. Dashed horizontal line indicate the Fermi energy.}
\label{Fig1}
\end{figure}

We begin by characterizing the electronic structure of bulk CrSBr and its evolution under surface electron doping via in situ potassium ($K$) deposition, providing a baseline for the subsequent analysis of the Gr/CrSBr heterostructure. The crystal and electronic properties of bulk CrSBr are summarized in Figure \ref{Fig1}. As illustrated in Fig. \ref{Fig1}(a), CrSBr crystallizes in an orthorhombic structure consisting of CrS bilayers encapsulated by anionic Br layers, which are stacked along the $c$-axis and separated by a vdW gap. Cleaving the crystal exposes a fresh $(a-b)$ plane, yielding a sharp, high-quality low-energy electron diffraction (LEED) pattern (Fig.~\ref{Fig1}(b)) that confirms the long-range orthorhombic symmetry of the surface.

The ARPES spectrum measured along high‑symmetry directions from this cleaved surface is presented in Fig.~\ref{Fig1}(c), revealing well‑defined dispersive states and a large band gap exceeding 1.65 eV. The valence‑band maximum (VBM) exhibits a parabolic dispersion, as highlighted in Fig.~\ref{Fig1}(d). These spectral features are well reproduced by our DFT calculations (Fig.~\ref{Fig1}(e)), with the exception of the absolute band‑gap value. Underestimation of band gap is typical for standard DFT treatment in correlated semiconductors and insulators. Incorporating on‑site electron correlation through a Hubbard \(U\) improves the calculated gap of CrSBr \cite{das2025surface, bianchi2023charge}, but the overall electronic structure including the gap magnitude and dispersion is more precisely captured by GW calculations \cite{smolenski2025large,bianchi2023paramagnetic,Watson_2024,wilson2021interlayer}. 

\begin{figure*}[ht!]
\centering
\includegraphics[width=14cm]{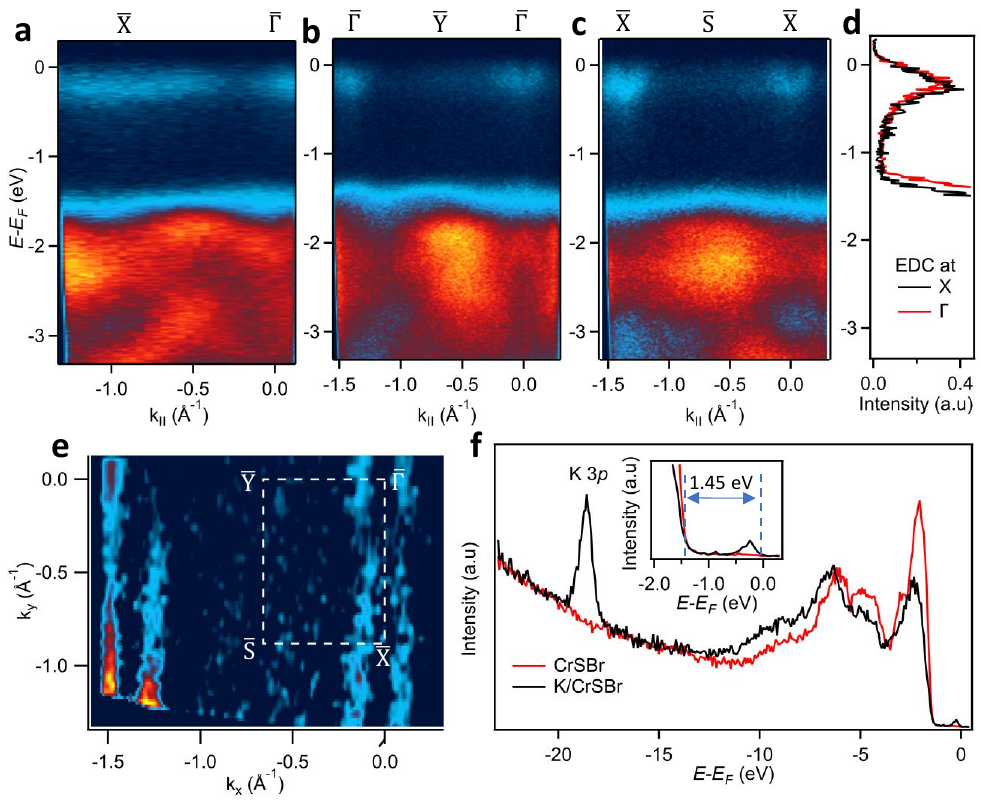}
\caption {Electronic structure of CrSBr after dosing of potassium on the surface. (a-c) ARPES spectra along $\bar{X}-\bar{\Gamma}$, $\bar{\Gamma}-\bar{Y}-\bar{\Gamma}$ and $\bar{X}-\bar{S}-\bar{X}$ high-symmetry paths, respectively using 40.8 eV photons. (d) EDC at ${\Gamma}$ and $\bar{X}$. (e) Fermi surface using 40.8 eV photons. (f) Integrated EDCs before and after potassium deposition. Inset shows zoomed-view close to the Fermi energy.}
\label{Fig2}
\end{figure*}

Interestingly, independent of the computational methods, CrSBr exhibits a robustly anisotropic conduction band topology, with a nearly flat dispersion along $\Gamma-X$ ($a$-axis) and highly dispersive behavior along $\Gamma-Y$ ($b$-axis). To experimentally access these states, we electron-doped the surface via potassium deposition, confirmed by the emergence of a clear K 3$p$ core‑level peak (Fig.~\ref{Fig2}(f)). ARPES spectra acquired along the high‑symmetry directions after dosing [Fig.~\ref{Fig2}(a-c)] reveal anisotropic electronic states: a nearly-flat conduction‑band segment along $\Gamma-X$ and highly dispersive electron pocket along $\Gamma-Y$, in good agreement with theory (Fig.~\ref{Fig1}(e)) and previous ARPES study on K/CrSBr system\cite{smolenski2025large}. This unique band topology manifests as a characteristic chain-like Fermi surface [Fig.~\ref{Fig2}(e)], suggesting quasi-1D electronic character. Notably, tuning flat bands toward the Fermi level is known to drive emergent phases such as unconventional magnetism, superconductivity, charge ordering, and heavy‑fermion behavior \cite{pakhira2025flat,li2019flat,kim2024tuning,cui2025theory,feuer2025charge,posey2024two}.

While the highly anisotropic shape of CrSBr crystal might suggest an anisotropic electronic structure, the microscopic origin of its quasi-1D character is non-trivial. Given that covalent Cr–S linkages extend along both the $a$ and $b$ axes within the structurally 2D layers, a more isotropic electronic structure might be expected. The finer details of the crystal structure suggest this quasi-1D behavior arises from the specific topology and hybridization of the Cr-S-Cr double-row chains extending along the $b$-axis. In this direction, Cr $3d$ orbitals (typically $d_{z^2}$) \cite{meng2026flat,ziebel2024crsbr,bianchi2023charge} are spatially oriented to point toward the sulfur atoms creates maximal overlap of the Cr $3d$ and S $3p$ orbitals facilitate a large hopping integral \cite{shi2025giant,ziebel2024crsbr}, yielding the steep, dispersive electron-like bands observed in Fig.~\ref{Fig2}. In contrast, the electronic coupling between adjacent chains along the $a$-axis is hindered by the intervening bromine (Br) atoms and a less favorable orbital geometry. This spatial decoupling effectively localizes charge carriers within individual 1D chains, leading to an exceptionally large effective mass and the observed flat-band character along the $\bar{\Gamma}-\bar{X}$ direction (Fig.~\ref{Fig2}(a)). The resulting chain-like Fermi surface [Fig.~\ref{Fig2}(e)] thus provides direct spectroscopic evidence of this emergent quasi-1D electronic character hosted within a structurally 2D lattice. Having established this anisotropic conduction‑band behavior, we next turn to the question of the band‑gap nature, which requires identifying the conduction‑band edge.

\begin{figure*}[ht!]
\centering
\includegraphics[width=16.5cm]{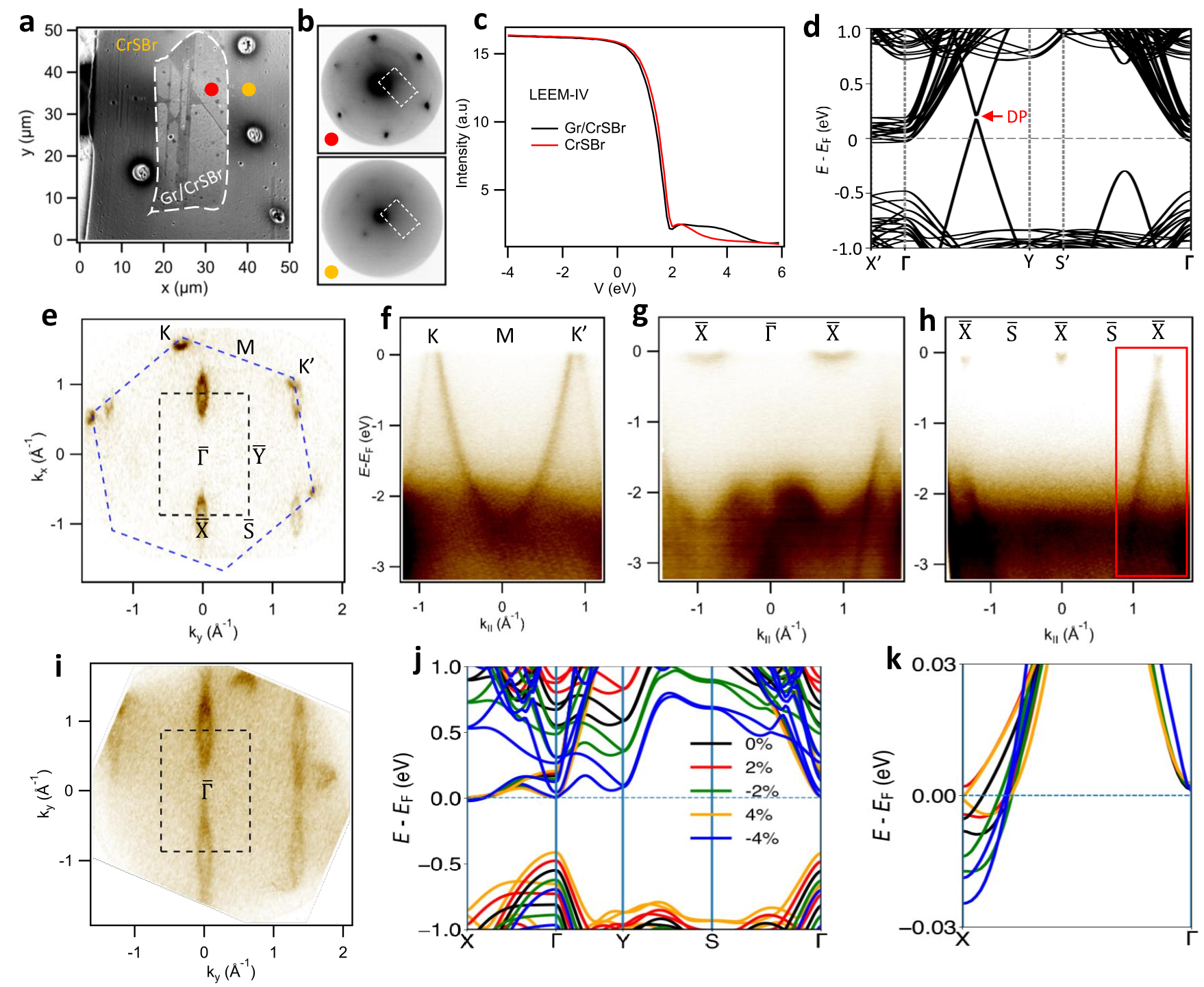}
\caption {Electronic structure of Gr/CrSBr heterostructure. (a) Mirror-mode LEEM image of the Gr/CrSBr. (b) Local diffraction patterns acquired at 50 eV from the Gr/CrSBr region (top) and the bare CrSBr region (bottom), corresponding to the circular markers in panel (a). Dashed-rectangle mark the reciprocal unit cell of CrSBr. (c) LEEM-IV from those two different regions. (d) DFT calculated band structure of the Gr/bilayer‑CrSBr heterostructure, with the Dirac point (DP) indicated. The X' and S' symmetry points are related to the folded CrSBr zone. (e) ARPES FS map of the heterostructure using 180 eV photon energy at 18 K. (f--h) ARPES spectra along high symmetry paths as indicated in figures. (i) ARPES FS map after the potassium deposition. We note that the K‑dosed measurements were performed on a separate heterostructure sample, in which the relative orientation between the Gr and CrSBr layers differed from that of the sample shown in panels (e–h). The dark and bright colors correspond to the maximum and minimum photoemission intensities, respectively. (j) Calculated electronic structure of bilayer CrSBr for various strain values as indicated (+ve tensile, --ve compressive). (k) Zoomed view around Fermi energy along $\Gamma-X$ from (j). In panel (j) and (k), Fermi level position was manually adjusted to mimic the experimental observation.}
\label{Fig3}
\end{figure*}

We already established that the VBM lies at $\Gamma$ (Fig.~\ref{Fig1}(d)). To determine the location of the conduction‑band minimum, we performed an energy‑distribution‑curve (EDC) analysis at the $\Gamma$ and $X$ points (Fig.~\ref{Fig2}(d)). The onset energies of CBM at these two momenta are nearly identical, with the minima at X appearing $\sim$ 25 meV lower than at $\Gamma$, consistent with our DFT results in Fig.~\ref{Fig1}(e). However, this small energy difference does not definitively establish whether CrSBr is a direct‑ or indirect‑gap semiconductor. Prior theoretical studies have shown that charge‑carrier injection in CrSBr can modify band alignment and orbital hybridization, leading to shifts in the relative positions of the band extrema \cite{bianchi2023charge}. Furthermore, depending on the computational approach, literature reports scatter between direct‑ and indirect‑gap character \cite{smolenski2025large,ziebel2024crsbr,telford2022coupling,shi2025giant,bianchi2023charge}. On the other hand, polarization-resolved photoluminescence (PL) results argued evidence of a direct gap \cite{wilson2021interlayer,telford2020layered}. Collectively, the results suggest that CrSBr sits near the proximity of a direct–indirect band‑gap crossover regimes.
Furthermore, after K dosing, the energy separation between the VBM and CBM decreases from its pristine value of $>$ 1.65 eV to $\sim$ 1.45 eV. This reduction likely originates from the carrier-induced band gap renormalization and the surface Stark effect \cite{smolenski2025large}. Similar charge‑transfer–driven gap narrowing has been reported for CrSBr flakes on Ag(111) and Au(111), where the fundamental gap evolves from direct on Au(111) to indirect on Ag(111) \cite{bianchi2023charge}.

We now turn to the Gr/CrSBr heterostructure. We first assess the structure of the Gr/CrSBr heterostructure using LEEM and micro‑LEED \cite{bauer2014surface}. LEEM imaging reveals large‑area graphene coverage on the CrSBr surface (Fig.~\ref{Fig3}(a)), with sharp contrast between bare CrSBr and Gr/CrSBr regions. Micro‑LEED patterns acquired from the Gr/CrSBr region (Fig.~\ref{Fig3}(b)) show intense  diffraction spots from graphene forming hexagon, accompanied by weaker reflections from the underlying CrSBr lattice; these CrSBr‑derived features become more clearly resolved in diffraction patterns collected from regions without graphene (Fig.~\ref{Fig3}(b), bottom). Additional insight into interfacial electronic reconstruction is provided by LEEM‑IV measurements. The LEEM‑IV curve from the Gr/CrSBr region (Fig.~\ref{Fig3}(c)) exhibits a pronounced enhancement in reflectivity near 3.5 eV compared to bare CrSBr, indicating increased local electron density possibly arising from the occupation of the CrSBr conduction band in the presence of graphene. 

These observations are qualitatively consistent with our DFT calculations of the Gr/CrSBr heterostructure, in which a single graphene layer is interfaced with a bilayer CrSBr slab (Fig.~\ref{Fig3}(d)). We emphasize that this minimal model is not intended to quantitatively reproduce the full heterostructure, particularly given the finite thickness of the CrSBr slab in our calculation.  Instead, the calculations are used to capture the essential features of the interfacial electronic structure, including band alignment, charge redistribution, and the emergence of conduction-band states near the Fermi level. Despite these limitations, the calculation reproduces the appearance of low-energy conduction-band states and the overall redistribution of spectral features near the Fermi level. Furthermore, we  observe a gapped Dirac point (with gap size $\sim$ 20 meV) in our calculations, indicative of strain‑induced sublattice symmetry breaking in graphene \cite{park2015band}.

The electronic structure of the heterostructure was further investigated using high-resolution micro‑ARPES. The Fermi‑surface map in Fig.~\ref{Fig3}(e) reveals elliptical pockets at the $\bar{X}$ point of the CrSBr Brillouin zone, together with the hole‑like graphene Dirac cones at $K$ and $K'$. \AK{From the relative orientation of the CrSBr and graphene BZs, we infer that the CrSBr $a$-axis ($\Gamma-X$) is roatated by \(19^\circ\) w.r.t the armchair direction of graphene ($\Gamma-M$).} Furthermore,The hole‑doped nature of graphene is clearly visible in the ARPES dispersion (Fig.~\ref{Fig3}(f)), from which the Dirac point is determined by extrapolating the linear bands to their crossing point, located 0.48 $\pm$ 0.03 eV above the Fermi level (Supporting Information (SI): Figure S1). The extracted Fermi velocity, \(v_F = 1.2\pm0.06 \times 10^6\) m/s, together with the measured Dirac‑point shift, yields a hole concentration of \(1.2\pm0.26\times 10^{13}\,\mathrm{cm^{-2}}\), obtained from the following equation  
\[
n = \frac{1}{\pi} \left( \frac{E_{\mathrm{Dirac}}}{\hbar v_F} \right)^2 .
\]
This hole density corresponds to the number of electrons expected to transfer from graphene into the interfacial CrSBr layers. To quantify the corresponding electron accumulation in CrSBr, we determined the FS area of the elliptical pockets at \(\bar{X}\) and used a Luttinger‑count analysis to extract the carrier density. The resulting value, \(1.5\pm0.2\times 10^{13}\,\mathrm{cm^{-2}}\), is in close agreement with the charge transferred from graphene. This corresponds to an occupation of $\sim$ 0.025 electrons per CrSBr unit cell.

The band dispersions of CrSBr along high‑symmetry directions are shown in (Fig.~\ref{Fig3}(g-h)). A highly anisotropic electron pocket is observed at the $\bar{X}$ point, forming the elliptical Fermi surface. By fitting the CrSBr conduction‑band dispersion to a parabola, we extract markedly anisotropic effective masses along the $\bar{\Gamma}-\bar{X}$ and $\bar{S}-\bar{X}$ (b-axis) directions, with values of 2.72 $\pm$ 0.12 $m_e$ and 0.18 $\pm$ 0.02 $m_e$, respectively (SI Figure S2). This pronounced mass anisotropy ($m_a/m_b\approx15$) provides a microscopic basis for the direction‑dependent optical and transport properties previously reported for this system \cite{wu2022quasi,rizzo2025engineering}. For example, in the case of uniaxial SPPs, this electronic anisotropy induce extreme directional anisotropy in the interface conductivity, which flattens the plasmonic isofrequency contours and forces the electromagnetic energy flow to align strictly with the axis of high carrier mobility. Notably, in contrast to potassium‑doped bulk CrSBr, where electron pockets appear at both $\bar{\Gamma}$ and $\bar{X}$ (Fig.~\ref{Fig2}(a-c)), the Gr/CrSBr heterostructure exhibits no detectable pocket at $\bar{\Gamma}$. To test whether this pocket can be restored, we performed potassium dosing on the heterostructure. As shown in (Fig.~\ref{Fig3}(i), the $\bar{X}$ pockets enlarge and eventually appear to meet as they extend towards $\bar{\Gamma}$, but no additional pocket emerges at $\bar{\Gamma}$. Further dosing only increase spectral broadening without altering the band topology. \AK{We note that the K-dosing measurements were performed on a seperate Gr/CrSBr sample, where the armchair direction of graphene were rotated by \(9^\circ\) w.r.t the $a$-axis of CrSBr.}

Moreover the above results indicate that the CBM of CrSBr lies at $\bar{X}$, in contrast to the potassium-doped case where CBM was nearly degenerate at both $\bar{X}$ and $\bar{\Gamma}$. \AK{Although the electronic structure of K‑dosed CrSBr is not a perfectly clean reference for pristine CrSBr, however, potassium adsorption is known in many semiconducting and insulating systems to reduce the band gap while leaving the overall band dispersion largely intact, aside from possible mass‑renormalization effects \cite{ni2024indium,spbv-1xg6,kim2015observation,kim2016determination}. As an alternative to K/CrSBr, a cleaner reference for the conduction‑band electronic structure could be intrinsically n‑doped bulk CrSBr, where recent ARPES measurements have revealed nearly degenerate conduction‑band valleys at $\bar{\Gamma}$ and $\bar{X}$ \cite{Biktagirov2025}, closely resembling the behavior observed in K‑dosed CrSBr.} All these together suggest a fundamental reconstruction of the CrSBr band structure and band alignment upon interfacing with graphene which cannot be explained just by considering charge transfer mechanism. The data further suggest an insulator‑to‑metal transition in the interfacial CrSBr layer when forming the Gr/CrSBr heterostructure, highlighting the ability of vdW interfaces to tune the intrinsic optoelectronic properties of 2D magnetic materials.

Interfacial strain and/or hybridization may drive the observed band‑structure reconstruction. Strain is expected at the interface because CrSBr (orthorhombic) and graphene (hexagonal) possess fundamentally different crystal symmetries, making lattice mismatch unavoidable. Our Raman spectroscopy further supports this idea where graphene G and 2D peaks shift to higher wavenumbers and develop low‑energy shoulders, consistent with peak splitting caused by reduced symmetry under uniaxial strain \cite{mohiuddin2009uniaxial} [SI: Figure S3]. As the main peaks exhibit a blue shift, the strain is likely uniaxial compressive, consistent with theoretical predictions \cite{rizzo2025engineering}. Moreover, such strain in graphene is expected to impose a corresponding strain field on the interfacial CrSBr layer. Complementary evidence comes from recent polarized photoluminescence (PL) measurements \cite{rizzo2025engineering}, which show reduced angular anisotropy in Gr/CrSBr compared to isolated few‑layer CrSBr, consistent with strain‑induced symmetry modification at the interface. To get deeper understanding about the strain effect, we performed band structure calculation using DFT under a range of compressive and tensile strain conditions. For simplicity, we modeled bilayer CrSBr and varied the applied strain. The resulting band structures (Fig.~\ref{Fig3}(j–k)) show that compressive strain drives the conduction‑band states at the \(X\) point to higher binding energy, whereas tensile strain shifts them in the opposite direction. In contrast, the electron pocket at \(\Gamma\) remains largely insensitive to strain. \AK{A similar strain‑dependent evolution of the electronic structure is observed for bulk CrSBr [SI: Figure S4].} This behavior indicates that compressive strain in the interfacial CrSBr layer naturally accounts for our experimental observations. \AK{Furthermore, calculations demonstrate that the conduction-band bandwidth is highly tunable via external strain, suggesting that lattice-induced deformation effectively modulates the underlying many-body interactions and the strength of electronic correlations in the system.} We note that the calculations presented here do not explicitly include the effects of electron doping, which is known to influence the band alignment and orbital hybridization in CrSBr \cite{bianchi2023charge}. However, as discussed earlier that charge transfer alone can not account for the observed behavior, we argue that interfacial compressive strain is the primary driver of the momentum-dependent band‑structure reconstruction in the Gr/CrSBr heterostructure. These results underscore that strain engineering, when combined with interfacial charge redistribution, offers a powerful route for tuning the band topology and electron correlations of CrSBr and should be applicable for other material systems.

\begin{figure}[h!]
\centering
\includegraphics[width=10cm]{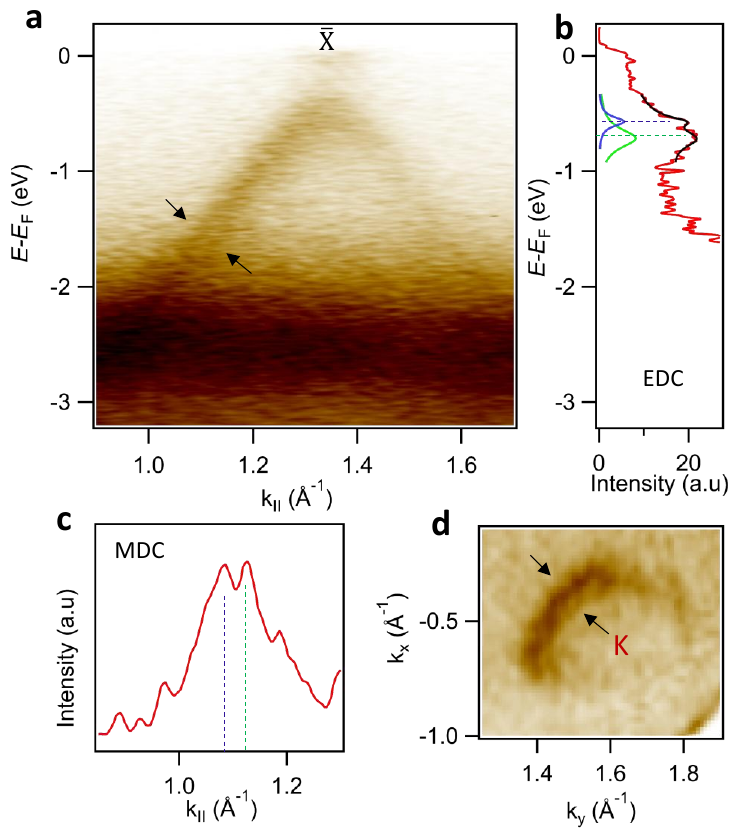}
\caption {Splitting of graphene Dirac bands. (a) Zoomed view of the ARPES spectrum enclosed by rectangle around $\bar{X}$ from Fig. 3(h). Splitting of graphene Dirac bands are marked by arrows. (b) Energy distribution curve (red) at $\bar{X}$ from (a). The EDC around the exchange splitting region were fitted using two Lorentzian peaks and the fitted envelope (black) is overlayed on EDC. (c) Momentum distribution curves at --1.65 eV from (a) shows clear splitting of Dirac band. (d) Constant‑energy contours around the K point reveal split pockets, indicated by arrows.}
\label{Fig4}
\end{figure}

\AK{We now turn to the possible spectroscopic signatures of proximity-induced magnetism in the graphene $\pi$-band. As shown in Fig.~\ref{Fig4}(a), the graphene $\pi$-band displays a splitting-like feature. An energy distribution curve (EDC) extracted at the $\bar{X}$ point (Fig.~\ref{Fig4}(b)) reveals two peak-like features centered $-0.65$ eV, with an estimated splitting of 120 $\pm$ 30 meV. This splitting is even more apparent in the momentum dispersion curve (MDC) cut at $-$1.65 eV as shown in Fig.~\ref{Fig4}(c), where the two momentum‑separated components are well resolved, producing the observed splitting of the K pocket (Fig.~\ref{Fig4}(d)). Similar features persist over a broader energy window, as seen in EDCs taken at multiple momenta (SI: Figures S5 and S6). The apparent separation increases at higher binding energies, suggesting its energy‑ and momentum‑dependency. Such behavior is qualitatively consistent with expectations from theoretical studies due to proximity induced magnetic exchange splitting in graphene \cite{yang2024electrostatically}. Furthermore, prior transport investigations have inferred the presence of exchange-induced band shifts in this system with an estimated exchange splitting of  $\sim 30$ meV \cite{yang2024electrostatically}. Notably, splitting estimated from ARPES spectrum significantly exceeds the value estimated from transport. A direct, one-to-one quantitative comparison between ARPES and transport is not straightforward because the two techniques probe different energy scales and distinct physical processes. Transport is sensitive only to states extremely close to the Fermi level, whereas ARPES shows that the magnitude of the band splitting varies strongly with momentum. Furthermore, the relative crystallographic orientation between graphene and CrSBr may also play a critical role, as theoretical models predict the magnetic proximity coupling strength to be acutely sensitive to the twist angle, with variations exceeding an order of magnitude (supplementary of Ref. \cite{yang2024electrostatically}). Additional enhancement could arise from the direct overlap between graphene states and the spin‑polarized CrSBr conduction band near the Fermi level (Fig.~\ref{Fig3}(e)), which may facilitate efficient interlayer hopping and hybridization \cite{cardoso2023strong}. Comparable magnitude of exchange splitting has recently been reported in graphene interfaced with other magnetic systems, including CrSe and Ni(111) \cite{wu2020large, zhang2018spin,sheverdyaeva2024spin}. However, its important to note that, although our observation of the split Dirac band is consistent with exchange splitting; alternative mechanisms such as lattice distortion, interfacial spin-orbit coupling, hybridization or sublattice symmetry breaking cannot be ruled out as possible origins of this behavior \cite{guinea2010energy,shakouri2013effect,nishidate2023gap}. Definitive identification of magnetic exchange would require spin-resolved ARPES which is beyond the scope of the present work. Furthermore, future studies exploring different relative orientations between graphene and CrSBr would be highly valuable for fully uncovering the mechanisms underlying proximity‑induced magnetic effects in this system, and such insights are likely to be broadly applicable to other van der Waals heterostructures as well.}

\section{Conclusions}
In summary, we present a comprehensive momentum‑resolved spectroscopic investigation of the Gr/CrSBr heterostructure, revealing how interfacial charge transfer, strain, and magnetic proximity collectively reshape the electronic landscape of this emerging 2D magnetic semiconductor system. LEEM and micro‑LEED measurements confirm large‑area, high‑quality graphene coverage on CrSBr lattice. Micro‑ARPES directly visualizes the resulting electronic reconstruction, showing strongly hole‑doped graphene, electron‑doped CrSBr with highly anisotropic 1D-like electronic states around $E_\mathrm{F}$. Furthermore, we show that the electronic structure of CrSBr undergoes substantial momentum-dependent renormalization within the heterostructure, exhibiting metallic behavior that cannot be explained solely by charge doping. We attribute this band‑structure reconstruction primarily to interfacial compressive strain in the CrSBr layer, consistent with our DFT calculations. Finally, we observe splitting-like behavior of graphene Dirac band, providing possible evidence of magnetic proximity in Gr/CrSBr. Moreover, our results establish Gr/CrSBr as model system for exploring the charge-transfer, strain-engineering, and magnetic proximity in low-dimensional quantum materials.

\section{Methods: Experimental and Theoretical details}

\subsection*{Sample preparation}

Single crystals of CrSBr were synthesized using a chemical vapor transport. The details of the CrSBr synthesis can be found in ref. \cite{bae2022exciton}.

Gr/CrSBr heterostructure samples were prepared using the Quantum Material Press (QPress) facility at the Center for Functional Nanomaterials (CFN), Brookhaven National Laboratory. Flakes of graphene and CrSBr were mechanically exfoliated onto O$_2$ plasma cleaned 90 nm SiO$_2$/Si and heavily doped Si substrates respectively in an argon filled glovebox. Single layer graphene (1LG) and few layer CrSBr ($\sim$ 30 nm) flakes were identified using optical contrast. While Raman spectroscopy (using 532 nm laser) was used to confirm the layer number of graphene from the intensity ratios of G and 2D peaks, atomic force microscopy (AFM) was used to confirm the thickness of CrSBr. 1LG/CrSBr/Si heterostructure was fabricated using dry transfer method. A transparent polydimethylsiloxane (PDMS) cube ($\sim 1\times1\times1$ mm$^3$) was covered by a thin polycarbonate (PC) polymer film attached to a glass slide served as the transfer stamp. First, 1LG was picked from the SiO$_2$ substrate at 110 $^\circ$C. The graphene flake on the PC film was then transferred onto the CrSBr flake on Si at an elevated temperature (170 $^\circ$C) by melting down the PC.The sample surface was subsequently washed with chloroform, acetone, and isopropyl alcohol to remove the melted PC polymer, and remaining residues were removed by gentle AFM cleaning.

For ARPES measurements, the completed samples were sealed in argon within the glovebox and transported to the beamline. Upon arrival, the seal was broken and the samples were rapidly loaded into the vacuum load‑lock, limiting air exposure to under one minute. The load‑lock reached a pressure of 4$\times$10\(^{-8}\) Torr within 45 minutes, after which the samples were transferred to a preparation chamber with a base pressure of 1$\times$10\(^{-10}\) Torr. Samples were annealed at 150 $^\circ$C for 2--3 hours to remove physisorbed surface contaminants, then loaded into the ARPES measurement chamber, operating at a base pressure of 2$\times$10\(^{-11}\) Torr.

\subsection*{Angle-resolved photoemission spectroscopy}

Bulk ARPES measurements were performed using a laboratory‑based system at the OASIS facility within the Condensed Matter Physics Department at Brookhaven National Laboratory (BNL). The setup is equipped with a Scienta R4000 electron analyzer and a VUV‑5k helium discharge lamp. Samples were cleaved {\it in situ} at room temperature under ultra‑high‑vacuum (UHV) conditions. During potassium deposition, samples were cooled to 30 K, and subsequent ARPES measurements were carried out at 200 K to mitigate charging effects.

Micro‑ARPES measurements on the Gr/CrSBr heterostructures were conducted at the Electron Spectro‑Microscopy (ESM) 21‑ID beamline of the National Synchrotron Light Source II at BNL \cite{rajapitamahuni2024electron}. The beamline is equipped with a Scienta DA30 analyzer and operates at a base pressure of \(2\times10^{-11}\) mbar, with a photon‑beam spot size of \(\sim 4\times6\) $\mu$m.  Measurements were performed at 18 K, with total energy and angular resolutions of approximately 20 meV and \(0.2^\circ\), respectively.

\subsection*{Low-energy electron microscopy}
After completing the ARPES measurements, the same sample was subsequently examined using low‑energy electron microscopy (LEEM). These measurements were performed in the aberration-corrected Elmitec LEEM III microscope at the XPEEM/LEEM endstation of the ESM beamline. The micro-LEED patterns have been acquired with the 1.5 um selected-area apertures. The LEEM IV spectra have been extracted from a set of LEEM images obtained with the field-of-view of 30 $\mu$m and landing energy of the electrons varied in the energy range of 10 eV with 0.1 eV energy step. Prior to LEEM/LEED measurements the sample has been degassed at \(160^\circ\)C in the UHV conditions for several hours.

\subsection*{Computational details}
Density functional theory (DFT) calculations were performed using the Quantum ESPRESSO package~\cite{QE}. The exchange–correlation functional was treated within the generalized gradient approximation using the Perdew–Burke–Ernzerhof (PBE) form~\cite{PBE}. No on-site Hubbard U correction was applied to the Cr-
3\textit{d} states, as the calculations are intended to provide qualitative insight into the band alignment and strain trends rather than quantitatively reproduce the correlated electronic structure. For bulk and bilayer CrSBr, Brillouin zone sampling was carried out using Monkhorst–Pack \textit{k}-point meshes of 
10$\times$10$\times$5 and 10$\times$10$\times$1, respectively. For graphene and graphene/CrSBr heterostructures, a reduced \textit{k}-point mesh of 
2$\times$3$\times$1 was employed due to the large supercell size.
The plane-wave kinetic energy cutoff for the wavefunctions was set to 40 Ry, with a corresponding charge density cutoff of 400 Ry. 

The graphene/CrSBr heterostructure was constructed by placing an 8$\times$2
 rectangular graphene supercell (32 carbon atoms) on top of a 5$\times$1 bilayer CrSBr supercell (20 Cr atoms per bilayer). In this configuration, the crystallographic \textbf{a} and \textbf{b} axes of CrSBr are aligned along the armchair and zigzag directions of graphene, respectively. A vacuum spacing of $\sim$ 20 \AA~ was introduced along the out-of-plane direction to eliminate spurious interactions between periodic images.
All structures were fully relaxed until the residual forces on each atom were less than $10^{-3}$  Ry/Bohr. Van der Waals interactions were included using the DFT-D3 correction scheme during structural relaxation. For the heterostructure calculations, the bottom CrSBr layer was kept fixed to mimic the bulk-like constraint from the underlying layers, while all other atoms were allowed to relax.
Spin-polarized calculations were performed with the CrSBr layers constrained in the A-type antiferromagnetic (AFM) configuration, consistent with the experimentally observed magnetic ground state

\section*{Acknowledgements}

This research used the resources of the Center for Functional Nanomaterials [Quantum Material Press (QPress)], and the National Synchrotron Light Source II (ESM beamline, 21-ID), which are U.S. DOE Office of Science Facilities, at Brookhaven National Laboratory under Contract No. DE-SC0012704. Synthesis work at Columbia was conducted as part of the Programmable Quantum Materials, an Energy Frontier Research Center, funded by the US Department of Energy (DOE), Office of Science, Basic Energy Sciences, under award no. DE-SC0019443.


\section*{Data availability}
The data that support the findings of this study are available from the corresponding author upon request.




\bibliography{refGr}

@Article{QE,
  author        = {Paolo Giannozzi and Stefano Baroni and Nicola Bonini et al},
  title         = {QUANTUM ESPRESSO: a modular and open-source software project for quantum simulations of materials},
  journal       = {Journal of Physics: Condensed Matter},
  year          = {2009},
  volume        = {21},
  number        = {39},
  pages         = {395502},
  __markedentry = {[aryal:6]},
  groups        = {From Paper},
  owner         = {niraj},
  timestamp     = {2016.12.20},
  url           = {http://stacks.iop.org/0953-8984/21/i=39/a=395502},
}

@Article{PBE,
  author    = {Perdew, John P. and Burke, Kieron and Ernzerhof, Matthias},
  title     = {Generalized Gradient Approximation Made Simple},
  journal   = {Phys. Rev. Lett.},
  year      = {1996},
  volume    = {77},
  pages     = {3865--3868},
  month     = {Oct},
  doi       = {10.1103/PhysRevLett.77.3865},
  issue     = {18},
  numpages  = {0},
  publisher = {American Physical Society},
  url       = {http://link.aps.org/doi/10.1103/PhysRevLett.77.3865},
}

@article{wilson2021interlayer,
  title={Interlayer electronic coupling on demand in a 2D magnetic semiconductor},
  author={Wilson, Nathan P and Lee, Kihong and Cenker, John and Xie, Kaichen and Dismukes, Avalon H and Telford, Evan J and Fonseca, Jordan and Sivakumar, Shivesh and Dean, Cory and Cao, Ting and others},
  journal={Nature Materials},
  volume={20},
  number={12},
  pages={1657--1662},
  year={2021},
  publisher={Nature Publishing Group UK London}
}

@article{telford2020layered,
  title={Layered antiferromagnetism induces large negative magnetoresistance in the van der Waals semiconductor CrSBr},
  author={Telford, Evan J and Dismukes, Avalon H and Lee, Kihong and Cheng, Minghao and Wieteska, Andrew and Bartholomew, Amymarie K and Chen, Yu-Sheng and Xu, Xiaodong and Pasupathy, Abhay N and Zhu, Xiaoyang and others},
  journal={Advanced Materials},
  volume={32},
  number={37},
  pages={2003240},
  year={2020},
  publisher={Wiley Online Library}
}

@article{bianchi2023charge,
  title={Charge transfer induced Lifshitz transition and magnetic symmetry breaking in ultrathin CrSBr crystals},
  author={Bianchi, Marco and Hsieh, Kimberly and Porat, Esben Juel and Dirnberger, Florian and Klein, Julian and Mosina, Kseniia and Sofer, Zdenek and Rudenko, Alexander N and Katsnelson, Mikhail I and Chen, Yong P and others},
  journal={Physical Review B},
  volume={108},
  number={19},
  pages={195410},
  year={2023},
  publisher={APS}
}

@article{wu2020large,
  title={Large exchange splitting in monolayer graphene magnetized by an antiferromagnet},
  author={Wu, Yingying and Yin, Gen and Pan, Lei and Grutter, Alexander J and Pan, Quanjun and Lee, Albert and Gilbert, Dustin A and Borchers, Julie A and Ratcliff, William and Li, Ang and others},
  journal={Nature Electronics},
  volume={3},
  number={10},
  pages={604--611},
  year={2020},
  publisher={Nature Publishing Group UK London}
}

@article{zhang2018spin,
  title={Spin-polarized semiconducting band structure of monolayer graphene on Ni (111)},
  author={Zhang, Yu and Sui, Xuelei and Ma, Dong-Lin and Bai, Ke-Ke and Duan, Wenhui and He, Lin},
  journal={Physical Review Applied},
  volume={10},
  number={5},
  pages={054043},
  year={2018},
  publisher={APS}
}

@article{yang2024electrostatically,
  title={Electrostatically controlled spin polarization in Graphene-CrSBr magnetic proximity heterostructures},
  author={Yang, Boxuan and Bhujel, Bibek and Chica, Daniel G and Telford, Evan J and Roy, Xavier and Ibrahim, Fatima and Chshiev, Mairbek and Cosset-Ch{\'e}neau, Maxen and Wees, Bart J van},
  journal={Nature Communications},
  volume={15},
  number={1},
  pages={4459},
  year={2024},
  publisher={Nature Publishing Group UK London}
}

@article{bae2022exciton,
  title={Exciton-coupled coherent magnons in a 2D semiconductor},
  author={Bae, Youn Jue and Wang, Jue and Scheie, Allen and Xu, Junwen and Chica, Daniel G and Diederich, Geoffrey M and Cenker, John and Ziebel, Michael E and Bai, Yusong and Ren, Haowen and others},
  journal={Nature},
  volume={609},
  number={7926},
  pages={282--286},
  year={2022},
  publisher={Nature Publishing Group UK London}
}

@article{rajapitamahuni2024electron,
  title={The electron spectro-microscopy (ESM) beamline at NSLS-II},
  author={Rajapitamahuni, A and Yilmaz, T and Kaznatcheev, K and Kundu, Asish K and Vescovo, E and Al-Mahboob, A and Sadowski, JT},
  journal={Synchrotron Radiation News},
  volume={37},
  number={4},
  pages={30--37},
  year={2024},
  publisher={Taylor \& Francis}
}

@article{rizzo2025engineering,
  title={Engineering anisotropic electrodynamics at the graphene/CrSBr interface},
  author={Rizzo, Daniel J and Seewald, Eric and Zhao, Fangzhou and Cox, Jordan and Xie, Kaichen and Vitalone, Rocco A and Ruta, Francesco L and Chica, Daniel G and Shao, Yinming and Shabani, Sara and others},
  journal={Nature communications},
  volume={16},
  number={1},
  pages={1853},
  year={2025},
  publisher={Nature Publishing Group UK London}
}

@article{ghiasi2021electrical,
  title={Electrical and thermal generation of spin currents by magnetic bilayer graphene},
  author={Ghiasi, Talieh S and Kaverzin, Alexey A and Dismukes, Avalon H and de Wal, Dennis K and Roy, Xavier and van Wees, Bart J},
  journal={Nature nanotechnology},
  volume={16},
  number={7},
  pages={788--794},
  year={2021},
  publisher={Nature Publishing Group UK London}
}

@book{bauer2014surface,
  title={Surface microscopy with low energy electrons},
  author={Bauer, Ernst},
  volume={23},
  year={2014},
  publisher={Springer-Verlag New York}
}

@article{smolenski2025large,
  title={Large exciton binding energy in a bulk van der Waals magnet from quasi-1D electronic localization},
  author={Smolenski, Shane and Wen, Ming and Li, Qiuyang and Downey, Eoghan and Alfrey, Adam and Liu, Wenhao and Kondusamy, Aswin LN and Bostwick, Aaron and Jozwiak, Chris and Rotenberg, Eli and others},
  journal={Nature Communications},
  volume={16},
  number={1},
  pages={1134},
  year={2025},
  publisher={Nature Publishing Group UK London}
}

@article{telford2022coupling,
  title={Coupling between magnetic order and charge transport in a two-dimensional magnetic semiconductor},
  author={Telford, Evan J and Dismukes, Avalon H and Dudley, Raymond L and Wiscons, Ren A and Lee, Kihong and Chica, Daniel G and Ziebel, Michael E and Han, Myung-Geun and Yu, Jessica and Shabani, Sara and others},
  journal={Nature Materials},
  volume={21},
  number={7},
  pages={754--760},
  year={2022},
  publisher={Nature Publishing Group UK London}
}

@article{mohiuddin2009uniaxial,
  title={Uniaxial strain in graphene by Raman spectroscopy: G peak splitting, Gr{\"u}neisen parameters, and sample orientation},
  author={Mohiuddin, TMG and Lombardo, Antonio and Nair, RR and Bonetti, A and Savini, G and Jalil, R and Bonini, Nicola and Basko, DM and Galiotis, C and Marzari, Nicola and others},
  journal={Physical Review B—Condensed Matter and Materials Physics},
  volume={79},
  number={20},
  pages={205433},
  year={2009},
  publisher={APS}
}

@article{ziebel2024crsbr,
  title={CrSBr: an air-stable, two-dimensional magnetic semiconductor},
  author={Ziebel, Michael E and Feuer, Margalit L and Cox, Jordan and Zhu, Xiaoyang and Dean, Cory R and Roy, Xavier},
  journal={Nano letters},
  volume={24},
  number={15},
  pages={4319--4329},
  year={2024},
  publisher={ACS Publications}
}

@article{rossi2023direct,
  title={Direct visualization of the charge transfer in a graphene/$\alpha$-RuCl3 heterostructure via angle-resolved photoemission spectroscopy},
  author={Rossi, Antonio and Johnson, Cameron and Balgley, Jesse and Thomas, John C and Francaviglia, Luca and Dettori, Riccardo and Schmid, Andreas K and Watanabe, Kenji and Taniguchi, Takashi and Cothrine, Matthew and others},
  journal={Nano Letters},
  volume={23},
  number={17},
  pages={8000--8005},
  year={2023},
  publisher={ACS Publications}
}

@article{lisi2021observation,
  title={Observation of flat bands in twisted bilayer graphene},
  author={Lisi, Simone and Lu, Xiaobo and Benschop, Tjerk and de Jong, Tobias A and Stepanov, Petr and Duran, Jose R and Margot, Florian and Cucchi, Ir{\`e}ne and Cappelli, Edoardo and Hunter, Andrew and others},
  journal={Nature Physics},
  volume={17},
  number={2},
  pages={189--193},
  year={2021},
  publisher={Nature Publishing Group UK London}
}

@article{xia2025superconductivity,
  title={Superconductivity in twisted bilayer WSe2},
  author={Xia, Yiyu and Han, Zhongdong and Watanabe, Kenji and Taniguchi, Takashi and Shan, Jie and Mak, Kin Fai},
  journal={Nature},
  volume={637},
  number={8047},
  pages={833--838},
  year={2025},
  publisher={Nature Publishing Group UK London}
}

@article{zhao2026second,
  title={Second-Harmonic Generation by Reconfigurable Lattice Engineering in Centrosymmetric Magnetic CrSBr},
  author={Zhao, Jinghan and Han, Xu and Bao, Xiaotian and Xue, Tongtong and Zeng, Xin and Tian, Yubo and Cai, Ningyi and Yan, Jiahao and Fu, Jianhui and Yue, Shuai and others},
  journal={Nano Letters},
  year={2026},
  publisher={ACS Publications}
}

@article{rassekh2026proximity,
  title={Proximity-induced spin-orbit torque in graphene on a trigonal CrSBr monolayer},
  author={Rassekh, Maedeh and Gmitra, Martin},
  journal={Physical Review B},
  volume={113},
  number={3},
  pages={035126},
  year={2026},
  publisher={APS}
}

@article{cao2018unconventional,
  title={Unconventional superconductivity in magic-angle graphene superlattices},
  author={Cao, Yuan and Fatemi, Valla and Fang, Shiang and Watanabe, Kenji and Taniguchi, Takashi and Kaxiras, Efthimios and Jarillo-Herrero, Pablo},
  journal={Nature},
  volume={556},
  number={7699},
  pages={43--50},
  year={2018},
  publisher={Nature Publishing Group UK London}
}

@article{li2025stacking,
  title={Stacking effects on magnetic, vibrational, and optical properties of CrSBr bilayers},
  author={Li, Huicong and Yang, Yali and Xia, Zhonghao and Wang, Yateng and Wei, Jiacheng and He, Jiangang and Wang, Rongming},
  journal={Physical Review B},
  volume={111},
  number={12},
  pages={125411},
  year={2025},
  publisher={APS}
}

@article{feuer2025charge,
  title={Charge density wave and ferromagnetism in intercalated CrSBr},
  author={Feuer, Margalit L and Thinel, Morgan and Huang, Xiong and Cui, Zhi-Hao and Shao, Yinming and Kundu, Asish K and Chica, Daniel G and Han, Myung-Geun and Pokratath, Rohan and Telford, Evan J and others},
  journal={Advanced Materials},
  volume={37},
  number={24},
  pages={2418066},
  year={2025},
  publisher={Wiley Online Library}
}

@article{Watson_2024,
 title={Giant exchange splitting in the electronic structure of A-type 2D antiferromagnet CrSBr}, 
 volume={8}, ISSN={2397-7132}, url={http://dx.doi.org/10.1038/s41699-024-00492-7}, DOI={10.1038/s41699-024-00492-7},
 number={1}, journal={npj 2D Materials and Applications}, 
 publisher={Springer Science and Business Media LLC},
 author={Watson, Matthew D. and Acharya, Swagata and Nunn, James E. and Nagireddy, Laxman and Pashov, Dimitar and Rösner, Malte and van Schilfgaarde, Mark and Wilson, Neil R. and Cacho, Cephise}, year={2024}, month=aug }

@article{kundu2020valence,
  title={Valence band electronic structure of the van der Waals ferromagnetic insulators: VI$_3$ and CrI$_3$},
  author={Kundu, Asish K and Liu, Yu and Petrovic, C and Valla, T},
  journal={Scientific reports},
  volume={10},
  number={1},
  pages={15602},
  year={2020},
  publisher={Nature Publishing Group UK London}
}

@article{marques2023interplay,
  title={Interplay between optical emission and magnetism in the van der Waals magnetic semiconductor CrSBr in the two-dimensional limit},
  author={Marques-Moros, Francisco and Boix-Constant, Carla and Ma{\~n}as-Valero, Samuel and Canet-Ferrer, Josep and Coronado, Eugenio},
  journal={ACS nano},
  volume={17},
  number={14},
  pages={13224--13231},
  year={2023},
  publisher={ACS Publications}
}

@article{klein2022control,
  title={Control of structure and spin texture in the van der Waals layered magnet CrSBr},
  author={Klein, J and Pham, T and Thomsen, JD and Curtis, JB and Denneulin, T and Lorke, M and Florian, M and Steinhoff, A and Wiscons, RA and Luxa, J and others},
  journal={Nature Communications},
  volume={13},
  number={1},
  pages={5420},
  year={2022},
  publisher={Nature Publishing Group UK London}
}

@article{das2025surface,
  title={Surface-dominated quantum-metric-induced nonlinear transport in the layered antiferromagnet CrSBr},
  author={Das, Kamal and Zhao, Yufei and Yan, Binghai},
  journal={Nano letters},
  volume={25},
  number={23},
  pages={9189--9196},
  year={2025},
  publisher={ACS Publications}
}

@article{bianchi2023paramagnetic,
  title={Paramagnetic electronic structure of CrSBr: Comparison between ab initio GW theory and angle-resolved photoemission spectroscopy},
  author={Bianchi, Marco and Acharya, Swagata and Dirnberger, Florian and Klein, Julian and Pashov, Dimitar and Mosina, Kseniia and Sofer, Zdenek and Rudenko, Alexander N and Katsnelson, Mikhail I and Van Schilfgaarde, Mark and others},
  journal={Physical Review B},
  volume={107},
  number={23},
  pages={235107},
  year={2023},
  publisher={APS}
}

@article{liu2018anisotropic,
  title={Anisotropic magnetocaloric effect in single crystals of CrI$_3$},
  author={Liu, Yu and Petrovic, Cedomir},
  journal={Physical Review B},
  volume={97},
  number={17},
  pages={174418},
  year={2018},
  publisher={APS}
}

@article{huidobro2016graphene,
  title={Graphene as a tunable anisotropic or isotropic plasmonic metasurface},
  author={Huidobro, Paloma A and Kraft, Matthias and Maier, Stefan A and Pendry, John B},
  journal={ACS nano},
  volume={10},
  number={5},
  pages={5499--5506},
  year={2016},
  publisher={ACS Publications}
}

@article{meng2026flat,
  title={Flat-band mechanism for strongly bound dark excitons in two-dimensional magnetic semiconductors},
  author={Meng, Zhong and Tian, Xirui and Wang, Aolei and Chu, Weibin and Zheng, Qijing and Zhao, Jin},
  journal={npj Computational Materials},
  year={2026},
  publisher={Nature Publishing Group UK London}
}

@article{shi2025giant,
  title={Giant magneto-exciton coupling in 2D van der Waals CrSBr},
  author={Shi, Jia and Wang, Dan and Jiang, Nai and Xin, Ziqian and Zheng, Houzhi and Shen, Chao and Zhang, Xinping and Liu, Xinfeng},
  journal={ACS nano},
  volume={19},
  number={33},
  pages={29977--29987},
  year={2025},
  publisher={ACS Publications}
}

@article{wu2022quasi,
  title={Quasi-1D electronic transport in a 2D magnetic semiconductor},
  author={Wu, Fan and Guti{\'e}rrez-Lezama, Ignacio and L{\'o}pez-Paz, Sara A and Gibertini, Marco and Watanabe, Kenji and Taniguchi, Takashi and von Rohr, Fabian O and Ubrig, Nicolas and Morpurgo, Alberto F},
  journal={Advanced Materials},
  volume={34},
  number={16},
  pages={2109759},
  year={2022},
  publisher={Wiley Online Library}
}

@article{park2015band,
  title={Band-gap opening in graphene: A reverse-engineering approach},
  author={Park, Joon-Suh and Choi, Hyoung Joon},
  journal={Physical Review B},
  volume={92},
  number={4},
  pages={045402},
  year={2015},
  publisher={APS}
}

@article{sheverdyaeva2024spin,
  title={Spin-Dependent $\pi$ $\pi$* Gap in Graphene on a Magnetic Substrate},
  author={Sheverdyaeva, PM and Bihlmayer, G and Cappelluti, E and Pacil{\'e}, D and Mazzola, F and Atodiresei, N and Jugovac, M and Grimaldi, I and Contini, G and Kundu, Asish K and others},
  journal={Physical Review Letters},
  volume={132},
  number={26},
  pages={266401},
  year={2024},
  publisher={APS}
}

@article{cardoso2023strong,
  title={Strong magnetic proximity effect in van der Waals heterostructures driven by direct hybridization},
  author={Cardoso, C and Costa, AT and MacDonald, AH and Fern{\'a}ndez-Rossier, J},
  journal={Physical Review B},
  volume={108},
  number={18},
  pages={184423},
  year={2023},
  publisher={APS}
}

@article{hallal2017tailoring,
  title={Tailoring magnetic insulator proximity effects in graphene: first-principles calculations},
  author={Hallal, Ali and Ibrahim, Fatima and Yang, Hongxin and Roche, Stephan and Chshiev, Mairbek},
  journal={2D Materials},
  volume={4},
  number={2},
  pages={025074},
  year={2017},
  publisher={IOP Publishing}
}

@article{zhao2025doping,
  title={Doping-induced magnetic phase transition enables all-electrical spin control in CrSBr},
  author={Zhao, Guorui and Zhao, Yibin and Zhang, Yu and Yang, Kunlin and Guo, Zejing and Liu, Jiaqi and Zhao, Tuoyu and Yan, Kun and Chen, Xiaobin and Li, Qi and others},
  journal={Nature Communications},
  year={2025},
  publisher={Nature Publishing Group UK London}
}

@article{yang2013proximity,
  title={Proximity Effects Induced in Graphene by Magnetic Insulators: First-Principles Calculations on Spin Filtering and Exchange-Splitting Gaps},
  author={Yang, Hong-Xin and Hallal, Ali and Terrade, D and Waintal, Xavier and Roche, Stephan and Chshiev, Mairbek},
  journal={Physical review letters},
  volume={110},
  number={4},
  pages={046603},
  year={2013},
  publisher={APS}
}

@article{kim2024tuning,
  title={Tuning the flat band with in-plane biaxial strain and the emergence of superconductivity in Ni$_3$Sn},
  author={Kim, Hye Jung and Kim, Min Jae and Lee, Jaekwang and Ok, Jong Mok and Kang, Chang-Jong},
  journal={Physical Review B},
  volume={110},
  number={2},
  pages={024504},
  year={2024},
  publisher={APS}
}

@article{pakhira2025flat,
  title={Flat-band tuning and emergent itinerant magnetism in Sr (Co$_{1-x}$Pd$_x$)$_2$As$_2$},
  author={Pakhira, Santanu and Lee, Yongbin and Kundu, Asish K and Islam, Farhan and Ning, Zhenhua and Smetana, Volodymyr and Mudring, Anja-Verena and Heitmann, Thomas and Vescovo, Elio and Ke, Liqin and others},
  journal={Proceedings of the National Academy of Sciences},
  volume={122},
  number={51},
  pages={e2519523122},
  year={2025},
  publisher={National Academy of Sciences}
}

@article{li2019flat,
  title={Flat-band magnetism and helical magnetic order in Ni-doped SrCo 2 As 2},
  author={Li, Yu and Liu, Zhonghao and Xu, Zhuang and Song, Yu and Huang, Yaobo and Shen, Dawei and Ma, Ni and Li, Ang and Chi, Songxue and Frontzek, Matthias and others},
  journal={Physical Review B},
  volume={100},
  number={9},
  pages={094446},
  year={2019},
  publisher={APS}
}

@article{posey2024two,
  title={Two-dimensional heavy fermions in the van der Waals metal CeSiI},
  author={Posey, Victoria A and Turkel, Simon and Rezaee, Mehdi and Devarakonda, Aravind and Kundu, Asish K and Ong, Chin Shen and Thinel, Morgan and Chica, Daniel G and Vitalone, Rocco A and Jing, Ran and others},
  journal={Nature},
  volume={625},
  number={7995},
  pages={483--488},
  year={2024},
  publisher={Nature Publishing Group UK London}
}

@article{cui2025theory,
  title={Theory of interaction-induced charge order in CrSBr},
  author={Cui, Zhi-Hao and Millis, Andrew J and Reichman, David R},
  journal={Physical Review B},
  volume={111},
  number={24},
  pages={245155},
  year={2025},
  publisher={APS}
}

@article{voloshina2026substrate,
  title={Substrate-induced magnetism in graphene: a minireview},
  author={Voloshina, Elena and Dedkov, Yuriy},
  journal={NPG Asia Materials},
  year={2026},
  publisher={Springer Japan Tokyo}
}

@article{guinea2010energy,
  title={Energy gaps and a zero-field quantum Hall effect in graphene by strain engineering},
  author={Guinea, Francisco and Katsnelson, Mikhail I and Geim, AK},
  journal={Nature Physics},
  volume={6},
  number={1},
  pages={30--33},
  year={2010},
  publisher={Nature Publishing Group UK London}
}

@article{shakouri2013effect,
  title={Effect of spin-orbit couplings in graphene with and without potential modulation},
  author={Shakouri, Kh and Masir, M Ramezani and Jellal, A and Choubabi, EB and Peeters, FM},
  journal={Physical Review B—Condensed Matter and Materials Physics},
  volume={88},
  number={11},
  pages={115408},
  year={2013},
  publisher={APS}
}

@article{nishidate2023gap,
  title={Gap opening at the Dirac point of graphene on Cu (111): Hybridization versus sublattice symmetry breaking},
  author={Nishidate, Kazume and Matsukawa, Michiaki and Hasegawa, Masayuki},
  journal={Surface Science},
  volume={728},
  pages={122196},
  year={2023},
  publisher={Elsevier}
}

@article{Biktagirov2025,
title={Intrinsic defects as a source of n-type conductivity in CrSBr},
volume={9},
ISSN={2397-7132},
DOI={10.1038/s41699-025-00640-7},
number={1}, 
journal={npj 2D Materials and Applications}, 
publisher={Springer Science and Business Media LLC}, 
author={Biktagirov, Timur and Schmidt, Wolf Gero and Schiller, Karl Jakob and Capra, Michele and Nitschke, Jonah Elias and Sternemann, Lasse and Arndt, Mira Sophie and Zamborlini, Giovanni and Isaeva, Anna and Cinchetti, Mirko}, 
year={2025},
month=Dec }

@article{spbv-1xg6,
  title = {Structurally Driven, Reversible Topological Phase Transition in a Distorted Square Net Material},
  author = {Yang, Xian P. and Hsu, Chia-Hsiu and Acharya, Gokul and Zhang, Junyi and Hossain, Md Shafayat and Cochran, Tyler A. and Neupane, Bimal and Cheng, Zi-Jia and Chhetri, Santosh Karki and Kim, Byunghoon and Gao, Shiyuan and Jiang, Yu-Xiao and Litskevich, Maksim and Wang, Jian and Wang, Yuanxi and Hu, Jin and Hasan, M. Zahid},
  journal = {Phys. Rev. Lett.},
  volume = {136},
  issue = {11},
  pages = {116603},
  numpages = {7},
  year = {2026},
  month = {Mar},
  publisher = {American Physical Society},
  doi = {10.1103/spbv-1xg6},
  url = {https://link.aps.org/doi/10.1103/spbv-1xg6}
}

@article{kim2015observation,
  title={Observation of tunable band gap and anisotropic Dirac semimetal state in black phosphorus},
  author={Kim, Jimin and Baik, Seung Su and Ryu, Sae Hee and Sohn, Yeongsup and Park, Soohyung and Park, Byeong-Gyu and Denlinger, Jonathan and Yi, Yeonjin and Choi, Hyoung Joon and Kim, Keun Su},
  journal={Science},
  volume={349},
  number={6249},
  pages={723--726},
  year={2015},
  publisher={American Association for the Advancement of Science}
}

@article{ni2024indium,
  title={Indium-Doped Crystals of SnSe$_2$},
  author={Ni, Danrui and Xu, Xianghan and Zhu, Zheyi and Ozbek, Yasemin and Mikšić Trontl, Vesna and Yang, Chen and Yang, Xiao and Louat, Alex and Cacho, Cephise and Ong, NP and others},
  journal={The Journal of Physical Chemistry C},
  volume={128},
  number={26},
  pages={11054--11062},
  year={2024},
  publisher={ACS Publications}
}

@article{kim2016determination,
  title={Determination of the band parameters of bulk 2H-MX$_2$ (M= Mo, W; X= S, Se) by angle-resolved photoemission spectroscopy},
  author={Kim, Beom Seo and Rhim, Jun-Won and Kim, Beomyoung and Kim, Changyoung and Park, Seung Ryong},
  journal={Scientific reports},
  volume={6},
  number={1},
  pages={36389},
  year={2016},
  publisher={Nature Publishing Group UK London}
}


\end{document}